# Tunable Circular Photogalvanic and Photovoltaic Effect in 2D Tellurium with Different Chirality

Chang Niu,[1,2] Shouyuan Huang,[2,3] Neil Ghosh,[2,3] Pukun Tan,[1,2] Mingyi Wang,[4] Wenzhuo Wu,[4] Xianfan Xu,[1,2,3]* and Peide D. Ye[1,2]*

[1]*Elmore Family School of Electrical and Computer Engineering, Purdue University, West Lafayette, IN 47907, United States.*

[2]*Birck Nanotechnology Center, Purdue University, West Lafayette, IN 47907, United States.*

[3]*School of Mechanical Engineering, Purdue University, West Lafayette, IN 47907, United States.*

[4]*School of Industrial Engineering, Purdue University, West Lafayette, IN 47907, United States.*

*Correspondence and requests for materials should be addressed to X.X. and P. D. Y. (xxu@ecn.purdue.edu and yep@purdue.edu)

Chirality arises from the asymmetry of matters, where two counterparts are the mirror image of each other. The interaction between circular-polarization light and quantum materials is enhanced in chiral space groups due to the structural chirality. Tellurium (Te) possesses the simplest chiral crystal structure, with Te atoms covalently bonded into a spiral atomic chain (left- or right-handed) with a periodicity of three. Here, we investigate the tunable circular photo-electric responses in 2D Te field-effect transistor with different chirality, including the longitudinal circular photogalvanic effect induced by the radial spin texture (electron-spin polarization parallel to the electron momentum direction) and the circular photovoltaic induced by the chiral crystal structure (helical Te atomic chains). Our work demonstrates the controllable manipulation of the chirality degree of freedom in materials.

Chirality is a fundamental property of particles, including electrons[1], photons, and phonons[2]. The chirality of photons describes the spin angular momentum direction parallel (right-handed) or antiparallel (left-handed) to the light propagation direction. This chirality degree of freedom in light interacts with the electron spin through angular momentum conservation. In solids, the spin-orbit-coupling-induced electron spin-polarization required by the crystal symmetry takes various spatial directions, at different momentum directions ***k***, called spin texture. The relation between electron spin texture in chiral materials and circular-polarization light are experimentally discussed using two-dimensional (2D) Tellurium (Te) as an example.

Te is an elemental narrow bandgap semiconductor with a chiral crystal structure[3]. Covalently bonded Te helical atomic chains are arranged in a hexagonal crystal lattice, as illustrated by **Figure 1(a)**. Te in a 2D form synthesized by the hydrothermal method[4] shows excellent electrical[5,6], thermal[7], and optical[8,9] properties. The chirality of the 2D Te flake is determined by the chirality of the left- or right-handed Te atomic chains, as shown in **Figure 1(b)**. Because of the mirror symmetry between two enantiomers, the asymmetry etchpits[10,11] formed by the hot sulfuric acid etching process can be used to characterize the chirality of 2D Te. The conduction band minimum of Te is located at the corner of the Brillouin zone ($H$ point). Due to the strong spin-orbit interaction[12] and the threefold screw symmetry, the spin-splitting conduction band cross at $H$ point and form a Weyl node[13,14], as shown in **Figure 1(c)**. The energy dispersion is the same between left- and right-handed Te. However, the topological charge of the Weyl node and the spin-polarization of the spin-splitting conduction bands are reversed[15]. 2D Te opens the door for realizing controllable chirality-based spin, electronic, or optical devices.

The spin texture in the Te conduction band is radial[5,16–18] (spin-polarization direction parallel to the electron momentum direction), as illustrated in **Figure 1(d)**. When the Fermi level is over the Weyl node energy, two Fermi surfaces (inner and outer) appear with opposite spin-polarization. The spin-polarization direction rotates once around $H$ point. The unique radial spin texture originates from the low symmetry of the crystal structure and the strong spin-orbit coupling (SOC) of Te atoms, reflected in the $k$ linear term of the conduction band Hamiltonian[14]: $H(k) = \frac{\hbar^2 k_\parallel^2}{2m_\parallel} + \frac{\hbar^2 k_\perp^2}{2m_\perp} + \hbar v_\parallel k_\parallel \sigma_z + \hbar v_\perp (k_x \sigma_x + k_y \sigma_y)$ (**Equation 1**). The Weyl-type SOC-induced[19] radial spin texture is different from that in the surface state of topological insulators[20–22] (**Figure 1(e)**, spin-polarization direction perpendicular to the electron momentum direction) or the Rashba SOC of semiconductor interfaces[23] (**Figure 1(f)**, spin-polarization direction perpendicular to the electron momentum direction, and two Fermi surfaces with opposite contribution). The left- and right-circular-polarized light with negative and positive angular momentum selectively excite electrons with opposite spin-polarizations according to angular momentum conservation. This chiral selection rule leads to the asymmetric populations of electrons with opposite spins, giving rise to a net spin-polarized electrical current.

In this paper, we systematically study the tunable circular optical electronic responses in 2D Te, including the longitudinal circular photogalvanic effect originating from the coupling between circular-polarized light and electron spins and the circular photovoltaic effect as a result of the Te chiral crystal structure. The circular photogalvanic and photovoltaic effects strongly depend on the chirality of the Te crystals, making 2D Te an ideal material system for developing chirality-dependent opto-electronic devices.

**Gate-tunable longitudinal circular photogalvanic effect in 2D Te**

The measurement setup is shown in **Figure 2(a)**. In our experiment, the 2D Te field-effect transistor (FET) is illuminated with a 633 nm He-Ne laser (~ 7 µm spot size) at room temperature. The polarization-dependent photocurrent was measured while rotating the quarter-wave plate (QWP) by an angle $\varphi$, which altered the laser polarization with a period of 180° from linear-polarization (LP) ($\varphi = 0°$) to right-circular-polarization (RCP) ($\varphi =$ 45°), to LP ($\varphi = 90°$), to left-circular-polarization (LCP) ($\varphi = 135°$), to LP ($\varphi = 180°$). $\theta$ is the angle between the incident light direction and x-y plane, $\alpha$ is the angle between the incident plane and x-axis (Te atomic chain direction and measured current direction). All the polarization-dependent photocurrent measured in this paper was under zero source-drain bias ($V_{ds} = 0\ V$). As illustrated in **Figure 2(b)**, the two Fermi surfaces have opposite spin-polarization directions. The contribution from two Fermi surfaces is opposite, resulting in net-zero spin current. When the circular-polarized light depopulates one branch of the 2D Te conduction band, however, a net spin-polarized current is generated due to the carrier concentration difference. Note that the spin-momentum configuration in 2D Te is different from that in topological insulators and Rashba SOC, the circular photogalvanic effect (CPGE) is longitudinal[24], where the generated photocurrent direction is parallel to the light incident direction.

**Figure 2(c)** is an optical image of a standard two-terminal 2D Te FET. The light is obliquely incident at $\theta = 45°$ along the current measurement direction from opposite directions ($\alpha = 0°\ and\ 180°$). A spatial-resolved photocurrent difference ($I_{RCP} - I_{LCP}$) mapping between RCP and LCP light is shown in **Figure 2(d)** ($\alpha = 0°$) and **Figure 2(e)** ($\alpha = 180°$). The photocurrent is quite uniform in the channel region of the device (which has an area of $30\ \mu m \times 75\ \mu m$). It has opposite signs under opposite incident light directions, indicating that the longitudinal CPGE originated from the radial spin texture of the 2D Te conduction band. The polarization-dependent photocurrent versus QWP angle $\varphi$ is described by: $I = D + C sin2\varphi + L_1 sin4\varphi + L_2 cos4\varphi$ (**Equation 2**)[20], where the $C$ term describes the circular-polarization sensitive photocurrent with a 180° period, $L_1$ and $L_2$ term depends on the linear-polarization of light with a 90° period, $D$ term is the polarization-independent photocurrent. **Figure 2(f)** shows the polarization-dependent photocurrent as a function of the QWP angle $\varphi$ fitted by **Equation 2** under normally ($\theta = 90°$) and obliquely incident light ($\theta = 45°$). The circular-polarization sensitive term under obliquely incident light is much larger compared to that in normally incident light, which is negligible, again suggesting that the CPGE is sensitive to the electron spin-polarization in 2D Te. Furthermore, the polarization-dependent photocurrent is also measured in two orthogonal directions (parallel/perpendicular to the Te atomic chain direction) under obliquely incident light at $\theta = 45°$ along $y$-direction ($\alpha = 90°$), as shown in **Figure S1**. The dependence of photocurrent on circular-polarization light is negligible when the current measurement direction is perpendicular to the incident plane. In conclusion, the circular-polarization-dependent photocurrent can only be generated along the obliquely incident direction due to the radial spin-momentum configuration induced by the Weyl node at the edge of the conduction band. The circular-polarization-dependent photocurrent $I_c$ extracted from **Equation 2** linearly depends on the power of the incident light, as shown in **Figure 2(g)**. The slope is calculated to be $10.9\ nA/mW$. We excluded the contribution from the circular photon drag effect caused by the linear momentum transfer due to the symmetry requirement of the Te crystal[25].

Due to the 2D nature of the Te flake, the chemical potential can be easily changed by the field effect. The observed polarization-sensitive photocurrent is tuned by applying a back gate voltage $V_{bg}$, as shown in **Figure 3(a)**. The transfer curve of the 2D Te FET at $V_{ds} = 0.05\ V$ (black curve **Figure 3(b)**) indicates that the device is operated in the n-type region. The back gate-dependent photocurrent fitted by **Equation 2** is shown in **Figure 3(b)**. Same tunable circular-polarization-dependent photocurrent is observed in another three different devices, as shown in **Figure S2**. The value of $C$ term is a function of the back gate voltage (electron density). We attribute this tunability to the different scattering rates of carriers under different gate voltages. Because of the unique chiral crystal structure of Te, the negative/positive Weyl nodes are separated in energy, resulting in a large topologically nontrivial window[19]. According to the theoretical prediction, the quantized circular-polarization-dependent saturation photocurrent generated from a single Weyl node is expressed as[26]: $I_c^{sat} = -\frac{4\pi\alpha e}{h}IC\tau$ (**Equation 3**), where $\alpha$ is the fine structure constant, $e$ is the elementary charge, $h$ is the Planck constant, $\tau$ is the electron scattering time, $I$ is the intensity of light, and $C = \pm 1$ is the topological charge of the Weyl node, which depends on the chirality of the 2D Te crystal, making 2D Te a promising candidate for investigating this exotic phenomenon. In our experiment setup, the photocurrent is measured at the steady-state condition, and the wavelength of the laser falls in the topologically nontrivial energy window. **Figure S3** shows the circular-polarization-dependent photocurrent $I_c^{sat}$ as a function of the scattering time estimated using the field-effect mobility $\mu_{FET} = \frac{e}{m^*}\tau$, where $m^*$ is the effective mass. The linear relation in scattering time and incident power indicates the possible topological origin of the CPGE in the 2D Te conduction band.

**Circular photovoltaic effect in 2D Te**

Besides the longitudinal CPGE, a circular photovoltaic effect (CPVE) is also observed in 2D Te FET. The device structure is shown in **Figure 4(a)**. 20 nm of atomic layer deposition (ALD) grown $Al_2O_3$ is used to dope the channel part of the Te flake into n-type[27]. However, the Te under the contact remains slightly p-doped due to the intrinsic accumulation layer[28], resulting a chemical potential difference near the contact. The band diagram of the device is shown in **Figure 4(b)**. The electron-hole pairs generated by the incident light are separated by the built-in electric field. The photocurrent has opposite direction at two different contacts. **Figure 4(c)** is a photocurrent mapping from normally incident linear-polarization light in a six-terminal Hall-bar 2D Te device. The positive and negative photocurrent generated in the source-drain direction at the edge of each contact indicates the photovoltaic effect in the 2D Te device.

This photovoltaic effect is also polarization-sensitive. **Figure 4(d)** is a photocurrent mapping of a two-terminal 2D Te FET under obliquely incident linear-polarization light.

The polarization-sensitive photocurrent as a function of the QWP angle $\varphi$ at two different contacts is shown in **Figure 4(e)**. Large opposite circular-polarization-dependent photocurrent $C$ dictated by the p-n junction direction at different contact is observed. A similar trend is observed in left and right contact when the absolute value of the photocurrent is plotted (**Figure 4(f)**). Unlike the CPGE observed previously, the CPVE photocurrent has a different amplitude in the same direction under LCP and RCP light. It does not depend on the incident angle $\theta$ $or$ $\alpha$ (**Figure S4**), indicating that the CPVE originated from the interaction between polarized light and the chiral crystal structure of 2D Te[29]. In other words, the electron-hole pairs generated by the LCP and RCP light is different. Using the chemical potential difference in 2D Te homojunction as an amplifier, this difference in LCP and RCP light is enlarged, providing a promising route for circular-polarization light detection.

**Chirality-dependent circular photogalvanic and photovoltaic effect**

As seen from previous descriptions, circular-polarized light is coupled with electron spins in Te, giving rise to the CPGE, whereas the difference in the population of the generated electron-hole pairs under LCP and RCP light results in the CPVE. Both effects are chirality-dependent, making 2D Te an ideal platform for chirality-based circular-polarization light detectors.

**Figure 5(a)** and **5(b)** are photocurrent mapping of left- and right-handed 2D Te devices under obliquely incident ($\theta = 45°$) linear-polarization light. The results are similar between two devices with negative photocurrent at left contact and positive photocurrent at right contact indicated by the gold dashed squares, because of the photovoltaic effect. **Figure 5(c)** and **5(d)** are the photocurrent difference ($I_{RCP} - I_{LCP}$) mapping of the same devices under obliquely incident ($\theta = 45°$) LCP and RCP light. Two different regions (near contacts and 2D Te channel) are clearly shown, separated by the gray dashed lines. The photocurrent difference generated by LCP and RCP light has opposite signs near two contacts dictated by the p-n junction direction, indicating the CPVE. The CPGE-induced photocurrent difference is nearly uniform across the 2D Te channel. Furthermore, the photocurrent difference is opposite in two enantiomers, indicating both effects are opposite in 2D Te with different chirality. The polarization-dependent photocurrent measured at the center of the devices as a function of the QWP angle $\varphi$ in left- (blue) and right-handed (red) 2D Te is plotted in **Figure 5(e)**, which clearly shows the opposite CPGE-induced $C$ term caused by 2D Te crystal chirality. This chirality-dependent photocurrent is also observed in other devices with electrons and holes conducting channels, shown in **Figure S5-S8**. Because electrons(holes) carry negative(positive) charges, the circular-polarization-dependent photocurrent direction reverses when the carrier type is switched. Because of this large chirality dependence of the CPGE and CPVE, it can be envisioned that highly sensitivity circular-polarization light detectors can be realized by simultaneously measuring the photogalvanic or the photovoltaic currents in left- and right-handed 2D Te.

In conclusion, we demonstrate two chirality-dependent tunable circular optical electronic responses in 2D Te FET, including the longitudinal circular photogalvanic effect, which is sensitive to the incident light direction, and the circular photovoltaic effect originated from the intrinsic chiral crystal structure. The strong and controllable coupling between 2D Te and circular-polarization light will enable the realization of chirality-based photodetection and the manipulation of spin and charge current using circular-polarization light.

**Materials and Methods**

**Growth of 2D Te flake.** The 2D Te flakes are grown by hydrothermal method. 0.09 g of $Na_2TeO_3$ and 0.5 g of polyvinylpyrrolidone (PVP) (Sigma-Aldrich) were dissolved in 33 ml double-distilled water under magnetic stirring to form a homogeneous solution. 3.33 ml of aqueous ammonia solution (25-28%, w/w%) and 1.67 ml of hydrazine hydrate (80%, w/w%) were added to the solution. The mixture was sealed in a 50 ml Teflon-lined stainless steel autoclave and heated at 180 °C for 30 hours before naturally cooling down to room temperature.

**Device fabrication.** Te flakes were transferred onto 90 nm $SiO_2$/Si substrate. The six-terminal Hall-bar devices were patterned using two step electron beam lithography 20/60 nm Ti/Au metal contacts were deposited by electron beam evaporation. 20 nm ALD $Al_2O_3$ was deposited onto the Te flakes at 200 °C using $(CH_3)_3Al$ (TMA) and $H_2O$ as precursors.

**Hot sulfuric acid etching of the 2D Te flakes.** The synthesized 2D Te flakes were transferred onto a 90 nm $SiO_2$/Si substrate. 2D Te flakes were cleaned following a DI water rinse and standard solvent cleaning process (acetone, methanol, and isopropanol). 2D Te flakes were etched in hot concentrated sulfuric acid at 100 °C for 5 min.

**Circular polarized photocurrent measurements.** A 633-nm He-Ne laser beam is focused using an objective (Mitutoyo 10x M Plan Apo) to a spot size ~ 7 μm and an incident angle at 45°. The laser spot is scanned along the channel of the device using a piezoelectric nano-positioner. A quarter-wave plate (QWP) is used to control the polarization of the beam. The photocurrent is measured using a DC source-meter (Keithley 2612A). All measurements are performed at zero source-drain bias and room temperature.

**Acknowledgements**

P.D.Y. was supported by Army Research Office under grant No. W911NF-15-1-0574. X.X. acknowledges the support by the National Science Foundation (CBET-2051525). W.W. was sponsored by the Army Research Office under Grant Number W911NF-20-1-0118. The synthesis of 2D Te was supported by NSF under grant no. CMMI-2046936. C.N. acknowledges valuable discussions with Xueji Wang and Yikang Chen. C.N. and N.G. acknowledge the technical support from Mauricio Segovia.

**Author Contributions**

P.D.Y. conceived and supervised the project. C.N. designed the experiments. C.N., S.H., and N.G. performed the photocurrent measurement under the supervision of X.X. M.W. synthesized the material under the supervision of W.W. C.N. and P.T. performed the hot sulfuric acid etching and fabricated the devices. C.N. and S.H. analyzed the data. C.N. wrote the manuscript and all the authors commented on it.

**Competing financial interests**

The authors declare no competing financial interest.

**Supporting Information**

Additional data and details for photocurrent measured at transverse incident light, the relation between saturation photocurrent and electron scattering time, gate voltage dependence of circular-polarization-dependent photocurrent, photovoltaic effect under normally incident light, chirality-dependent photocurrent mapping of hole and electron conducting channels.

Supporting Figures S1-S8.

**Corresponding Author**

*Xianfan Xu (E-mail: xxu@ecn.purdue.edu) and Peide D. Ye (E-mail: yep@purdue.edu)

**Figures**

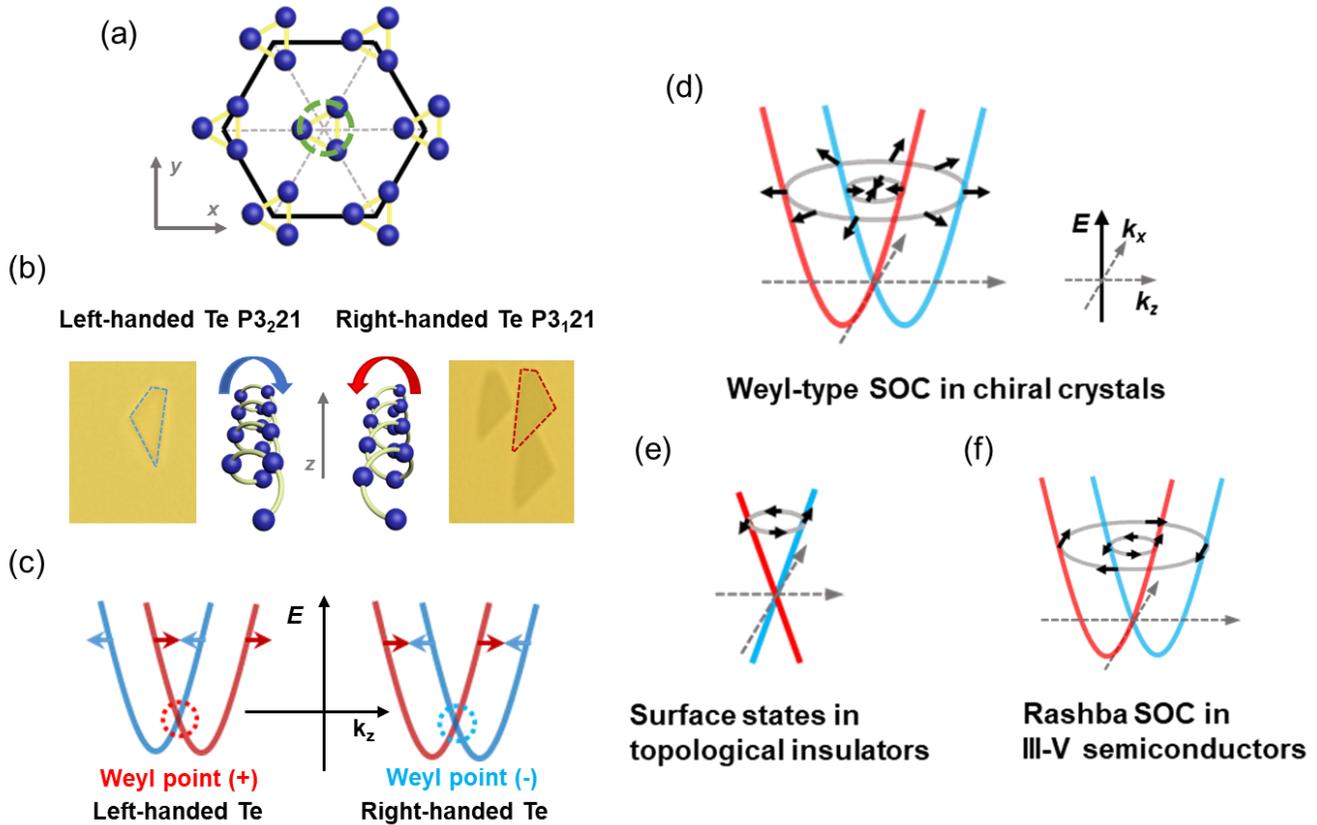

**Figure 1. Electron band structure and spin texture of left- and right-handed Te. (a)** The crystal structure of Te. **(b)** The helical Te atomic chains determine the chirality of the Te crystal. Different etch pits are observed in 2D Te after hot sulfuric acid etching. **(c)** The energy dispersion of left- and right-handed Te is the same. The spin polarization and the Weyl node charge are opposite between the two enantiomers due to the mirror symmetry. **(d-f)** The spin-texture in various materials: **(d)** Weyl-type SOC in chiral crystals (spin-polarization parallel to momentum direction), **(e)** surface states in topological insulators (spin-polarization perpendicular to momentum direction), and **(f)** Rashba SOC in III-V semiconductors (spin-polarization perpendicular to momentum direction).

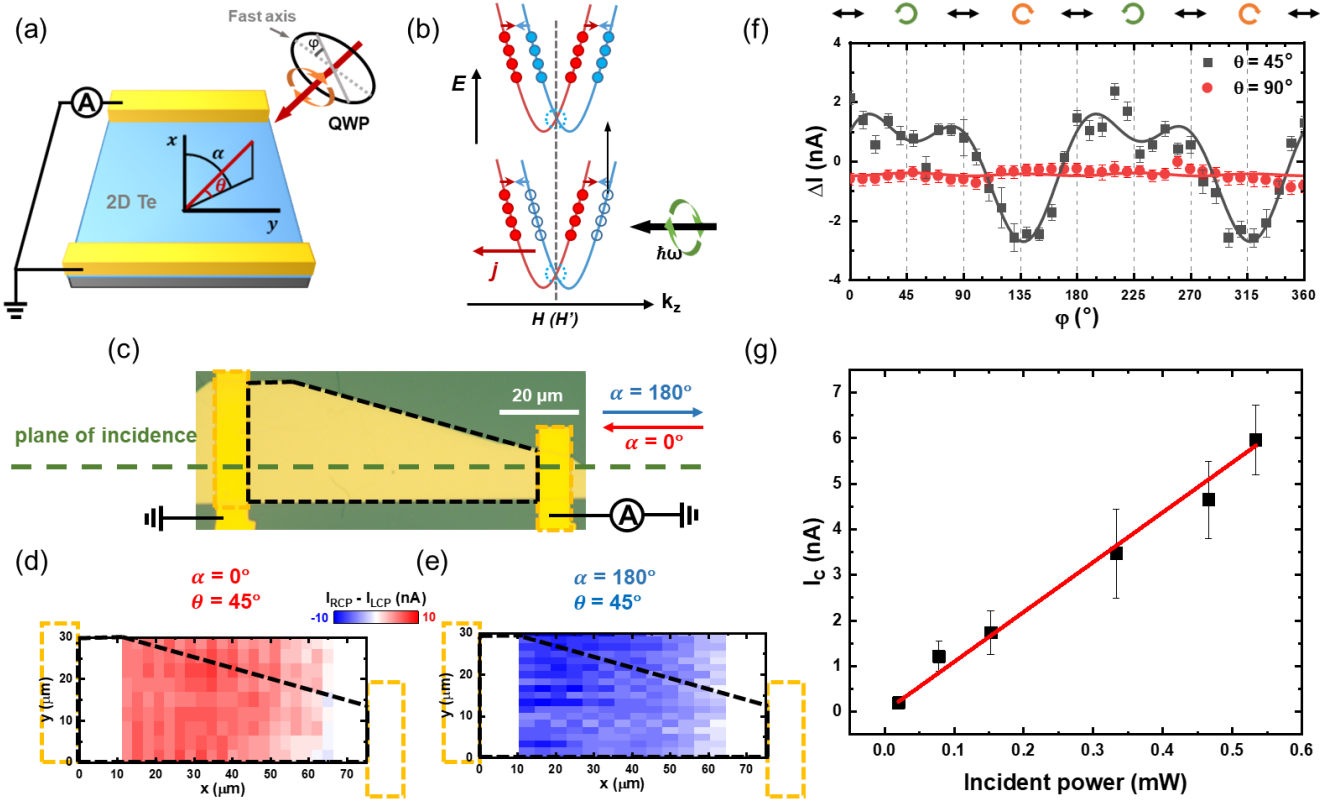

**Figure 2. Circular photogalvanic effect in 2D Te.** (**a**) Schematic of the circular-polarized photocurrent measurement setup. (**b**) The spin-polarized electrical current induced by the circular-polarized light. (**c**) Optical image of a two-terminal 2D Te field-effect transistor. The green dashed line indicates the plane of incidence. Spatially resolved photocurrent mapping of the difference between right circular-polarized (RCP) light and left circular-polarized light (LCP) at different incident conditions: (**d**) $\alpha = 0°, \theta = 45°$, (**e**) $\alpha = 180°, \theta = 45°$. (**f**) Photocurrent as a function of the quarter wave plate angle $\varphi$ at normally incident light ($\theta = 90°$) and obliquely incident light ($\theta = 45°$). (**g**) Circular-polarized light-dependent photocurrent $I_c$ as a function of laser intensity $I$.

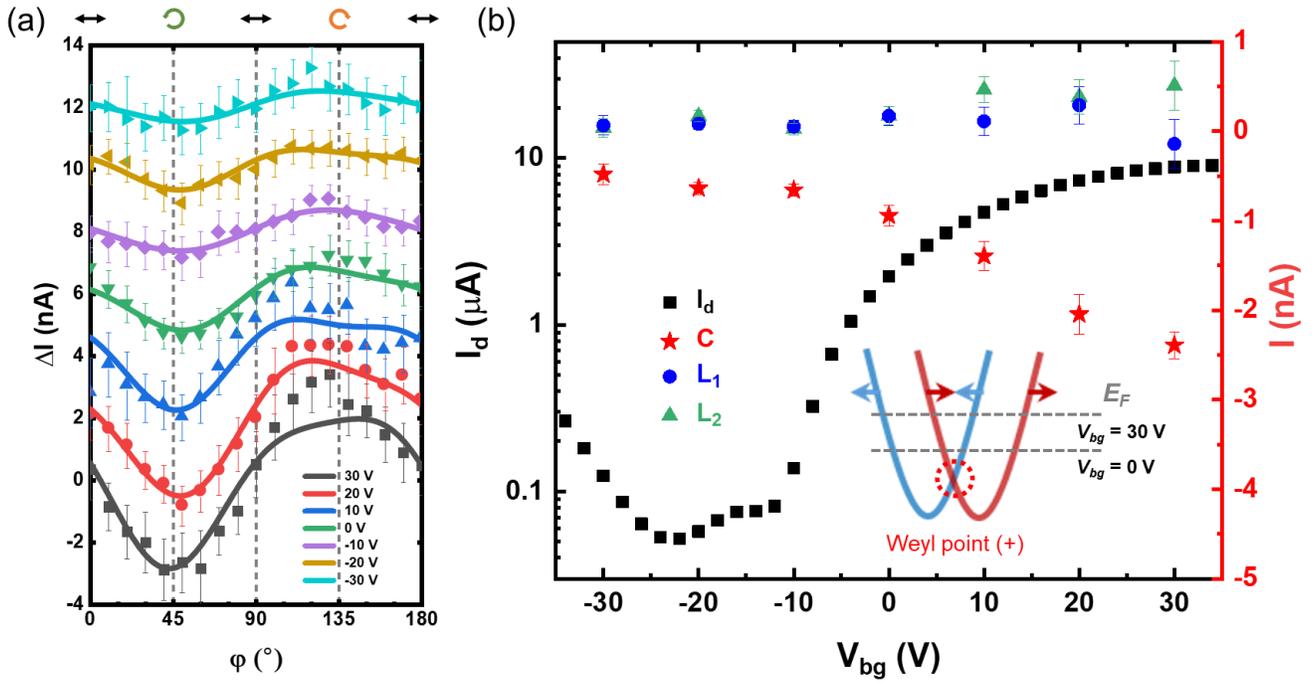

**Figure 3. Gate-tunable circular photogalvanic effect in 2D Te.** **(a)** The polarization-dependent photocurrent ΔI as a function of the quarter wave plate angle $\varphi$ at different back gate voltages. **(b)** Output characteristic (black) of a 2D Te field-effect transistor. Fit results calculated from **(a)**: C (red), $L_1$ (blue), and $L_2$ (green) as a function of the back gate voltage $V_{bg}$. **Inset:** the band structure of 2D Te. The back gate voltage tunes the Fermi level indicated by the dashed line.

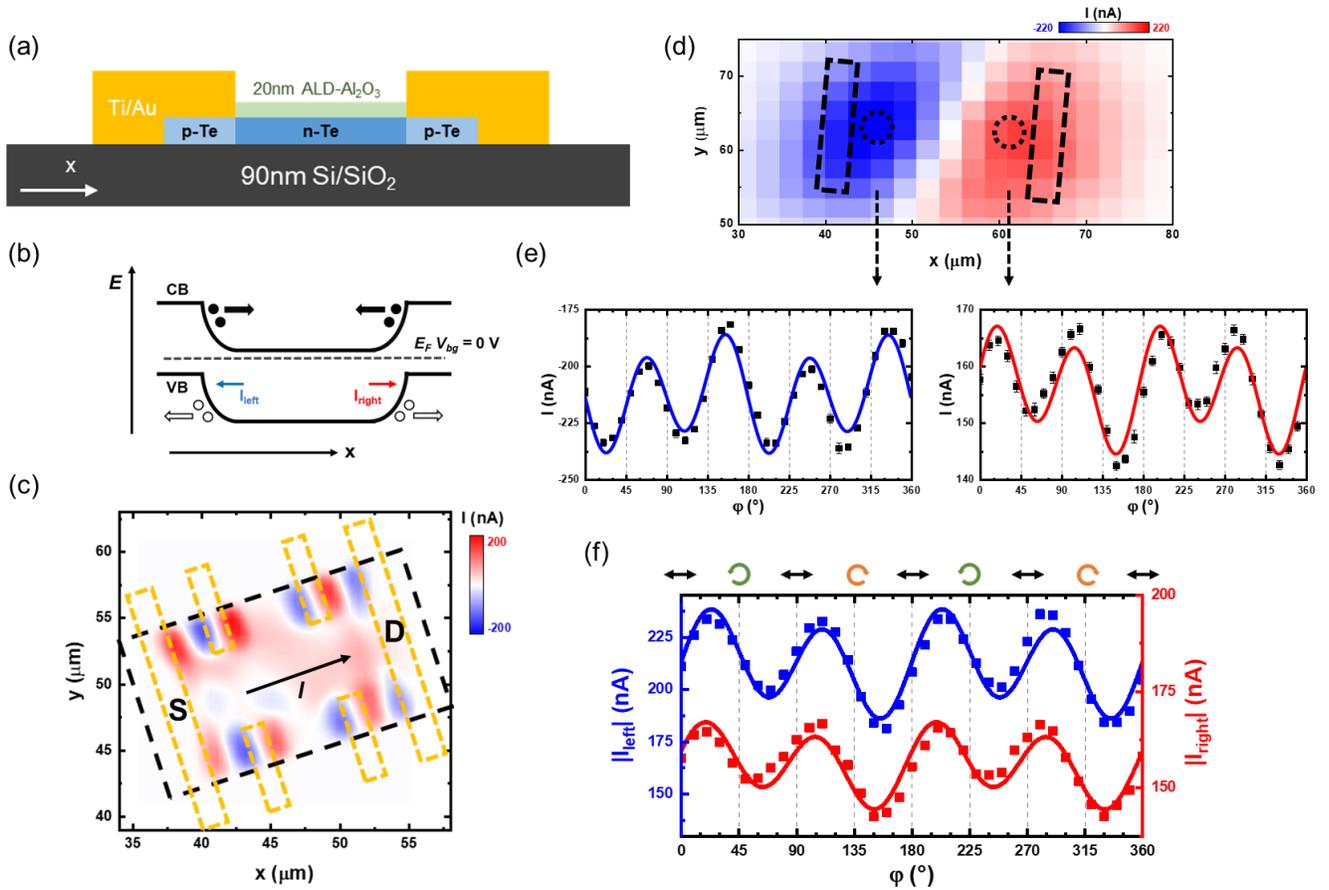

**Figure 4. Circular photovoltaic effect in 2D Te.** (**a**) Schematic of the ALD doped 2D Te device structure. A p-n junction is formed at the edge of the contact due to the different chemical potential of the 2D Te. (**b**) Band diagram of the 2D Te field-effect transistor. The direction of the photocurrent is dictated by the built-in field of the p-n junction. (**c**) Photocurrent mapping of a six-terminal Hall bar device under normally incident linearly polarized light. (**d**) Photocurrent mapping of a two-terminal Hall bar device under oblique incident linearly polarized light. (**e**) Photocurrent as a function of the quarter wave plate angle $\varphi$ at left and right contact. (**f**) The absolute value of the photocurrent is the same as shown in (**e**).

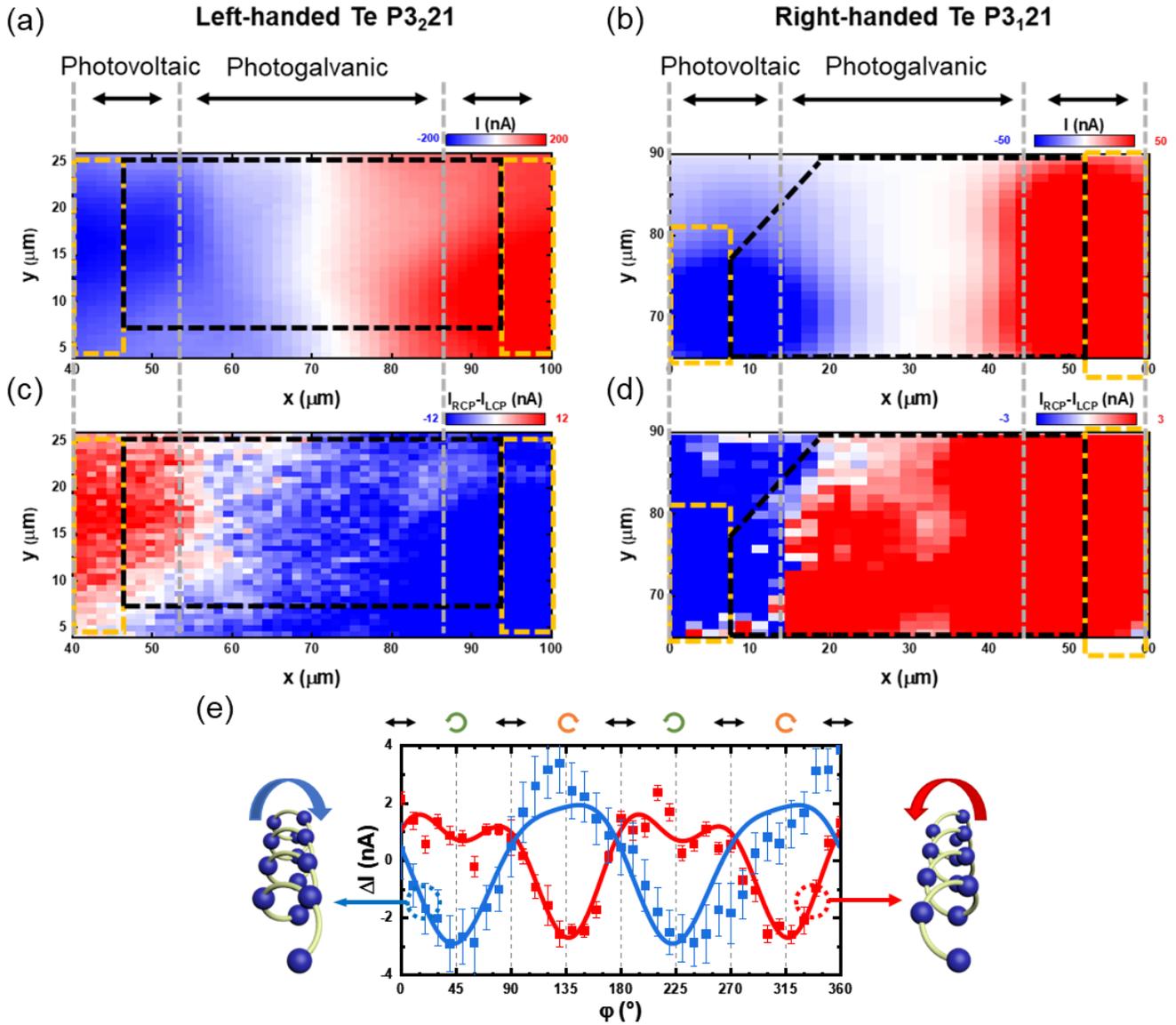

**Figure 5. Chirality-dependent circular photogalvanic and photovoltaic effect in 2D Te. (a)** Photocurrent mapping of a left-handed 2D Te device under oblique incident linearly polarized light. **(b)** Photocurrent mapping of a right-handed 2D Te device under the same condition. **(c)** Photocurrent difference mapping of a left-handed 2D Te device between oblique incident RCP and LCP light. **(d)** Photocurrent difference mapping of a right-handed 2D Te device under the same condition. **(e)** Polarization-dependent photocurrent as a function of the quarter wave plate angle $\varphi$ in left- and right-handed 2D Te.

**TOC Graphic**

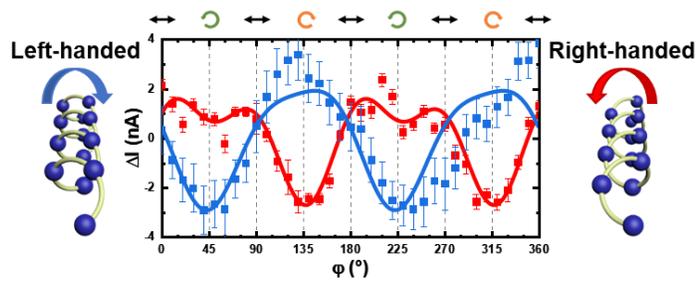

Supporting Information for:

# Tunable Circular Photogalvanic and Photovoltaic Effect in 2D Tellurium with Different Chirality

Chang Niu,[1,2] Shouyuan Huang,[2,3] Neil Ghosh,[2,3] Pukun Tan,[1,2] Mingyi Wang,[4] Wenzhuo Wu,[4] Xianfan Xu,[1,2,3]* and Peide D. Ye[1,2]*

[1]*Elmore Family School of Electrical and Computer Engineering, Purdue University, West Lafayette, IN 47907, United States.*

[2]*Birck Nanotechnology Center, Purdue University, West Lafayette, IN 47907, United States.*

[3]*School of Mechanical Engineering, Purdue University, West Lafayette, IN 47907, United States.*

[4]*School of Industrial Engineering, Purdue University, West Lafayette, IN 47907, United States.*

*Correspondence and requests for materials should be addressed to X.X. and P. D. Y. (xxu@ecn.purdue.edu and yep@purdue.edu)

# 1. Polarization-dependent photocurrent in transverse incident $\alpha = 90°$

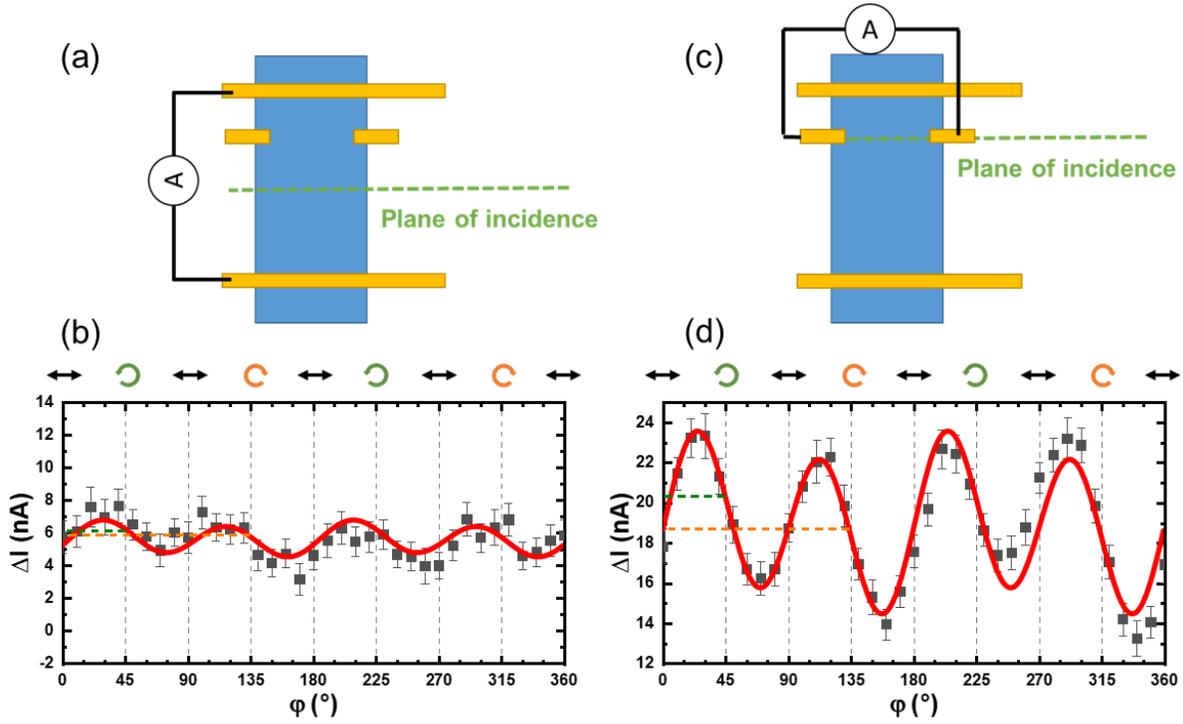

**Figure S1 (a)** Schematic of photocurrent measurement in $z$-direction under obliquely incident light at $\theta = 45°$ along $x$-direction ($\alpha = 90°$). **(b)** Polarization-dependent photocurrent as a function of QWP angle $\varphi$ using the setup in (a). Negligible circular-polarization dependent photocurrent is observed. **(c)** Schematic of photocurrent measurement in $x$-direction under obliquely incident light at $\theta = 45°$ along $x$-direction ($\alpha = 90°$). **(d)** Polarization-dependent photocurrent as a function of QWP angle $\varphi$ using the setup in (c). Large circular-polarization dependent photocurrent is observed, indicating a radial spin texture in 2D Te.

## 2. Tunable polarization-dependent photocurrent as a function of QWP angle $\varphi$

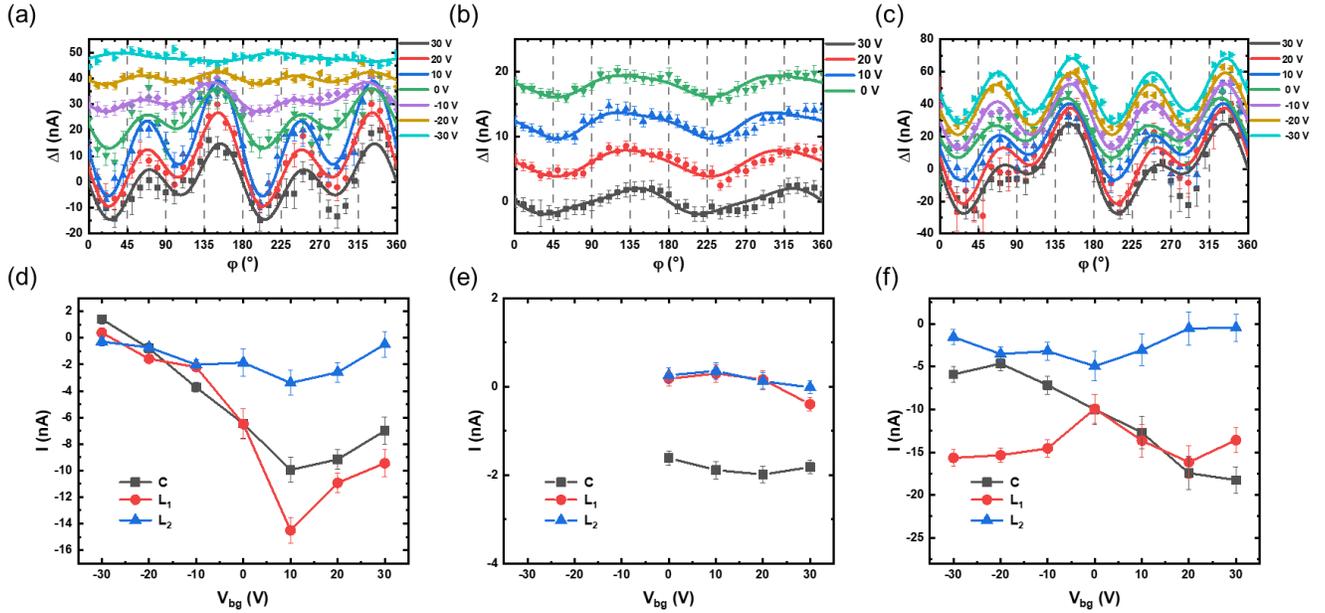

**Figure S2 (a-c)** Polarization-dependent photocurrent as a function of the QWP angle $\varphi$ at different back gate voltages in three different devices under obliquely incident light at $\theta = 45°$. The solid lines are curves fitting by **Equation 2**. **(d-f)** The back gate voltage dependence of fitting parameters $C$, $L_1$, $L_2$ extracted using (a-c) in three different devices. The circular-polarization-dependent photocurrent is tuned by the back gate voltages.

## 3. Circular-polarization-dependent photocurrent as a function of electron scattering time

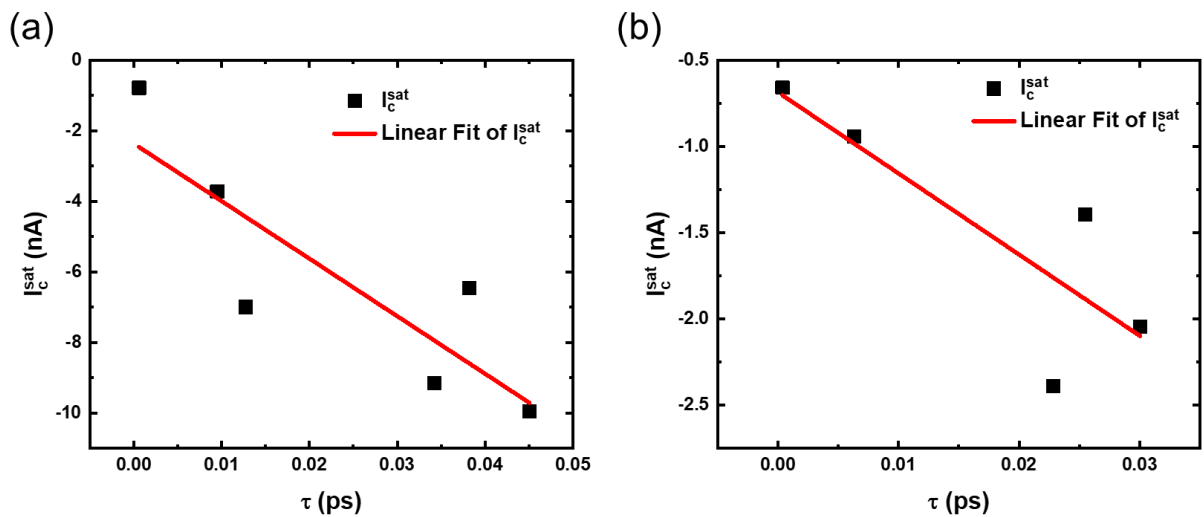

**Figure S3 (a, b)** Circular-polarization-dependent saturation photocurrent $I_c^{sat}$ as a function of the electron scattering time extracted from the field-effect mobility. Red lines are linear fit of the data in two different devices.

## 4. Circular photovoltaic effect under normally incident light.

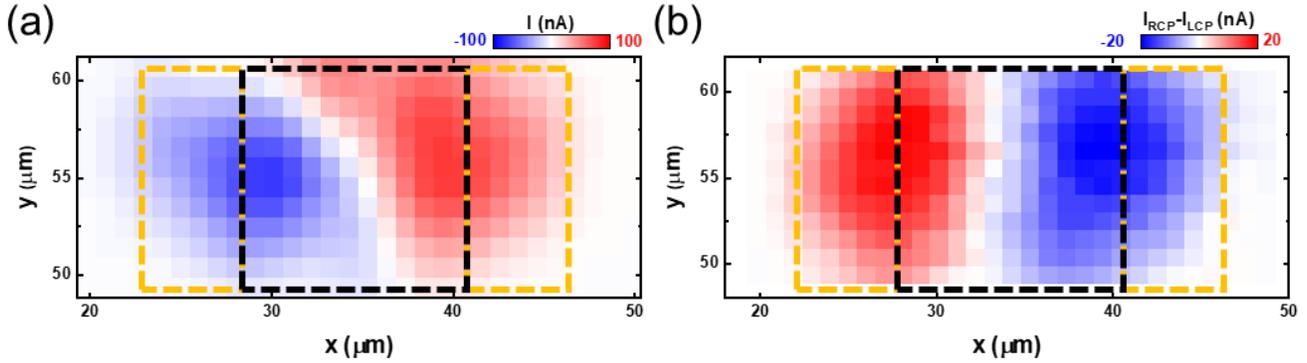

**Figure S4 (a)** Photocurrent mapping of a two-terminal 2D Te FET under normally incident linear-polarization light. **(b)** Photocurrent difference mapping between LCP and RCP normally incident light. Large circular-dependent photocurrent is observed with opposite directions at two contacts, indicating the CPVE is not sensitive to the light incident direction.

# 5. Circular-polarization-dependent photocurrent mapping in left-handed 2D Te holes

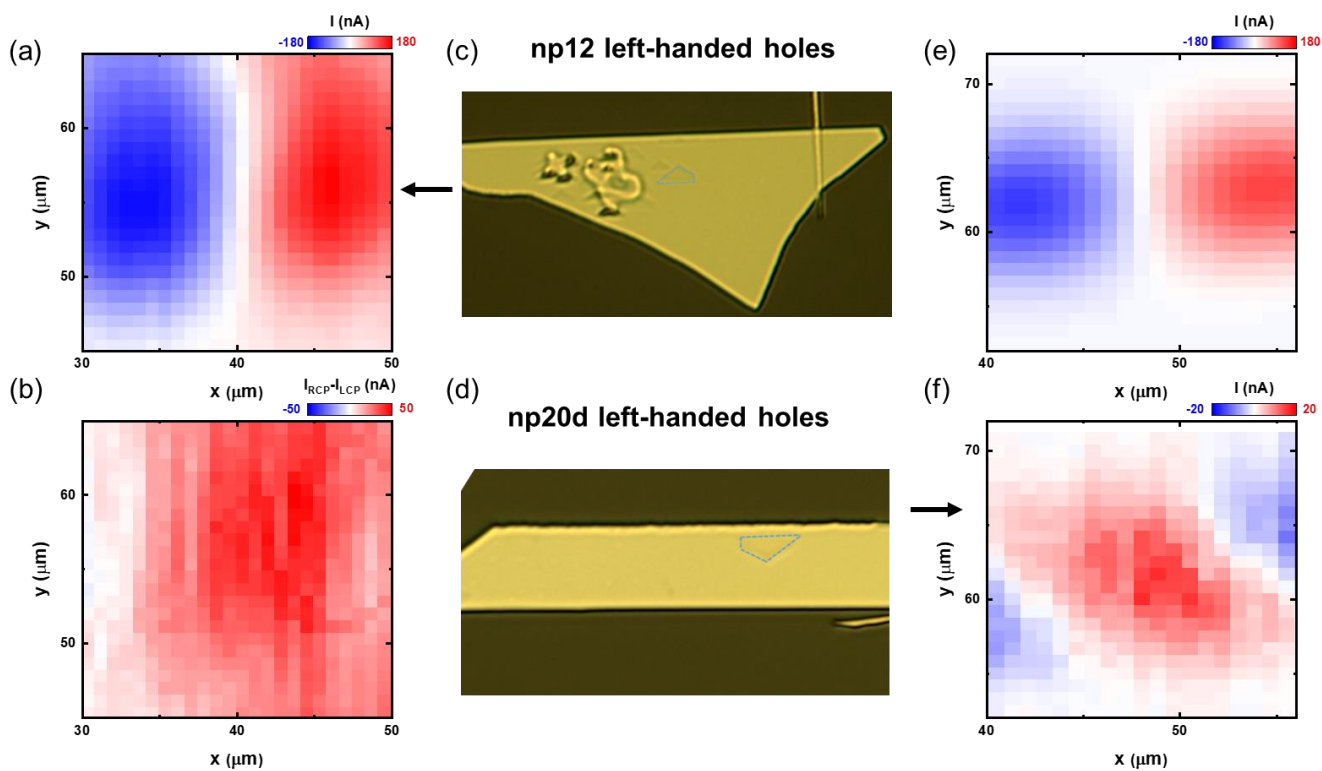

**Figure S5 (a, e)** Photocurrent mapping of two different devices under obliquely incident linear-polarization light. **(b, f)** Photocurrent difference mapping between LCP and RCP of two devices. **(c, d)** Optical image of two left-handed Te flakes with etchpits identified. The channel is hole conducting.

# 6. Circular-polarization-dependent photocurrent mapping in left-handed 2D Te electrons

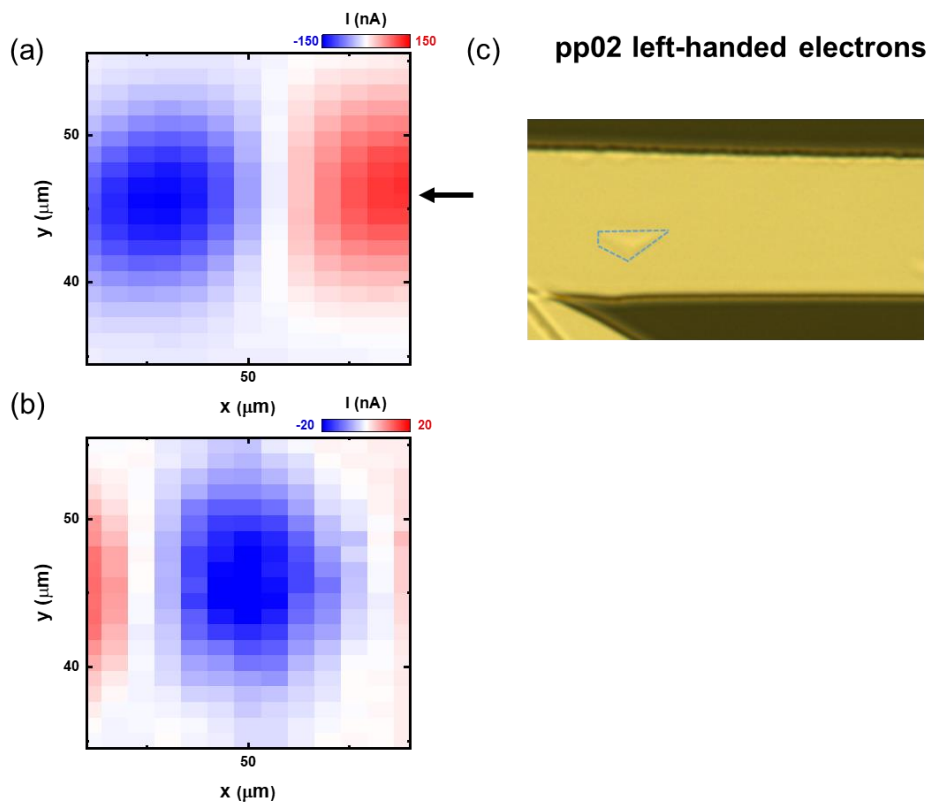

**Figure S6 (a)** Photocurrent mapping under obliquely incident linear-polarization light. **(b)** Photocurrent difference mapping between LCP and RCP. **(c)** Optical image of a left-handed Te flake with etchpits identified. The channel is electron conducting.

# 7. Circular-polarization-dependent photocurrent mapping in right-handed 2D Te holes

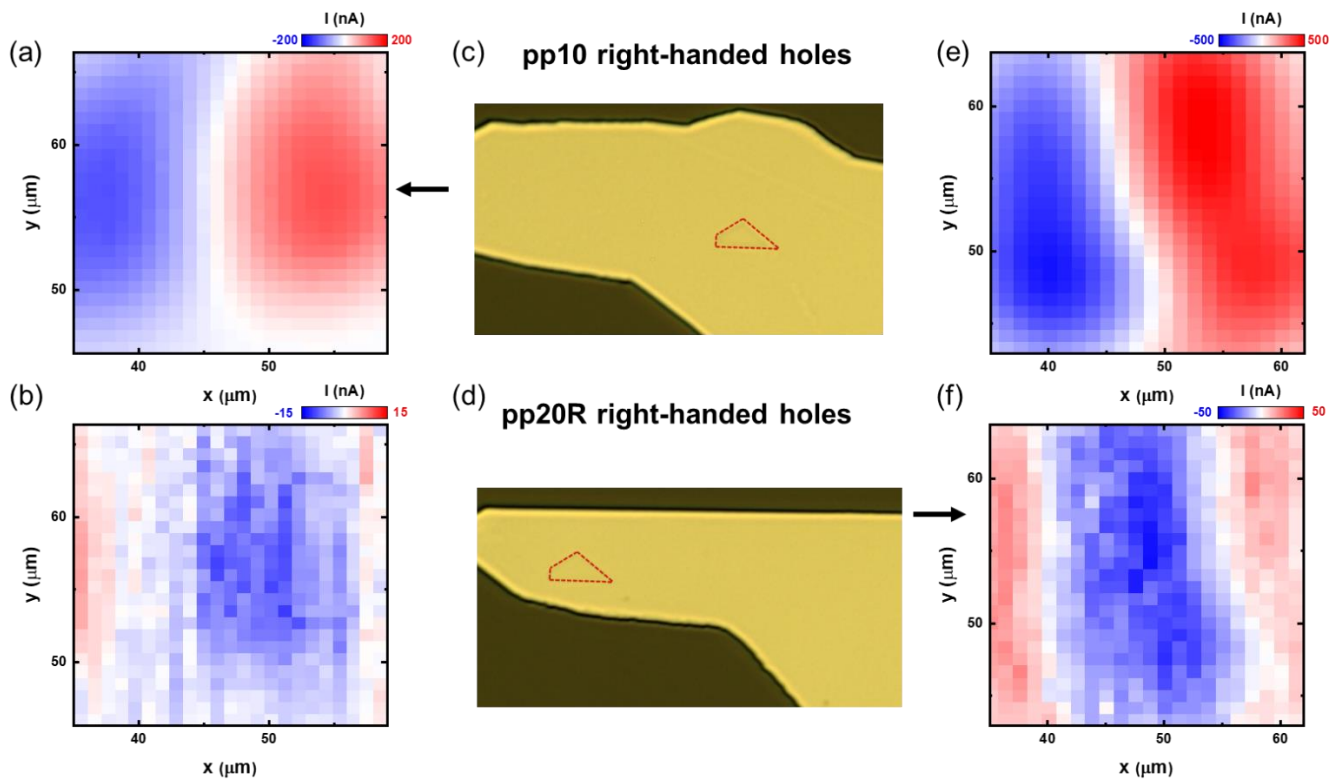

**Figure S7 (a, e)** Photocurrent mapping of two different devices under obliquely incident linear-polarization light. **(b, f)** Photocurrent difference mapping between LCP and RCP of two devices. **(c, d)** Optical image of two right-handed Te flakes with etchpits identified. The channel is hole conducting.

# 8. Circular-polarization-dependent photocurrent mapping in right-handed 2D Te electrons

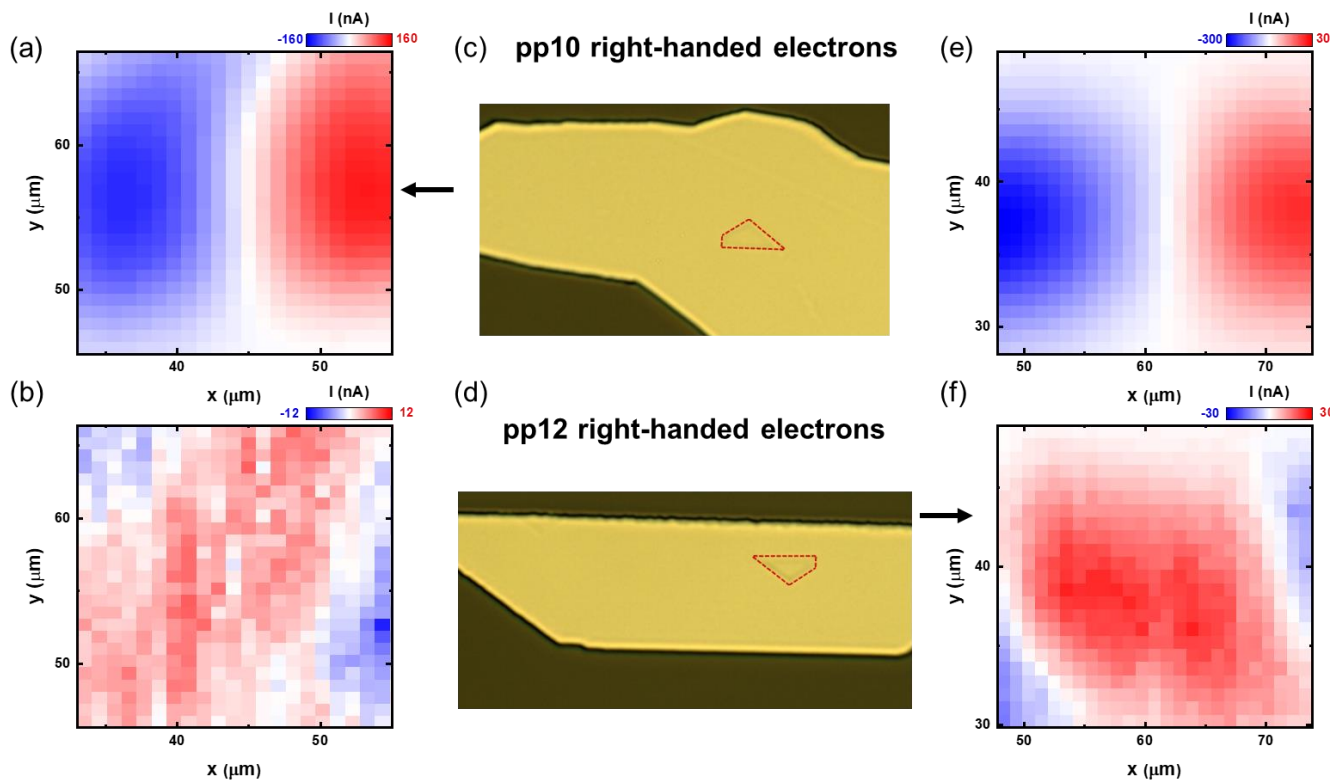

**Figure S8 (a, e)** Photocurrent mapping of two different devices under obliquely incident linear-polarization light. **(b, f)** Photocurrent difference mapping between LCP and RCP of two devices. **(c, d)** Optical image of two right-handed Te flakes with etchpits identified. The channel is electron conducting.